\documentclass[twocolumn,showpacs,amssymb,nofootinbib,floatfix]{revtex4-1}
\usepackage{epsfig,color}
\usepackage{slashed}
\newcommand{\be}{\begin{eqnarray}}
\newcommand{\ee}{\end{eqnarray}}
\newcommand{\order}[1]{ \mathcal{O} \left( #1 \right) }

  \newcommand{\lqcd}{\Lambda_{\mathrm{QCD}}}

\begin{document} \hbadness=10000
\topmargin -0.8cm\oddsidemargin = -0.7cm\evensidemargin = -0.7cm
\title{Photon Signals from Quarkyonic Matter}
\author{Giorgio Torrieri$^{1,2}$, Sascha Vogel$^{1,3}$,Bj{\o}rn B\"auchle$^{1}$}
\thanks{torrieri@fias.uni-frankfurt.de, svogel@fias.uni-frankfurt.de, baeuchle@th.physik.uni-frankfurt.de}
\affiliation{$^{1}$FIAS, Goethe Universit\"at, Ruth-Moufang-Str. 1, 60438 Frankfurt am Main, Germany \\
$^{2}$ Pupin Physics Laboratory, Columbia University, 538 West 120$^{th}$ Street, New York,
NY 10027, USA\\
 $^{3}$SUBATECH,
Laboratoire de Physique Subatomique et des Technologies Associ\'ees \\
University of Nantes - IN2P3/CNRS - Ecole des Mines de Nantes \\
4 rue Alfred Kastler, F-44072 Nantes Cedex 03, France
}
\begin{abstract}
We calculate the Bremsstrahlung photon spectrum emitted from dynamically evolving quarkyonic matter, and compare this spectrum with that of a high chemical potential quark-gluon plasma as well as to a hadron gas.  We find that the transverse momentum distribution and the harmonic coefficient is markedly different in the three cases.
The transverse momentum distribution of quarkyonic matter can be fit with an exponential, but is markedly steeper than the distribution expected for the quark-gluon plasma or a hadron gas, even at the lower temperatures expected in the critical point region.   The quarkyonic elliptic flow coefficient fluctuates randomly from event to event, and within the same event at different transverse momenta.  The latter effect, which can be explained by the shape of quark wavefunctions within quarkyonic matter, might be considered as a quarkyonic matter signature, provided initial temperature is low enough that the quarkyonic regime dominates over deconfinement effects, and the reaction-plane flow can be separated from the fluctuating component.
\end{abstract}
\maketitle
The study of nuclear matter at moderate ($T \sim 0-180$ MeV) temperature and large baryochemical potential ($\mu_Q = \mu_B/3 \sim \lqcd=250$ MeV) has recently enjoyed new vigorous theoretical \cite{review} and experimental interest.

From the experimental side, this is due to the start of programs specifically aimed at exploring lower energy collisions with the latest detector technology \cite{low1,low2,low3,low4}.
This regime presents both potential for very interesting physics and unique challenges, since an unambiguous lattice exploration is lacking \cite{latmu1,latmu2,latmu3},  effective field theory \cite{stephanov} gives ambiguous predictions, and the system remains profoundly non-perturbative \cite{casher1,casher2}.

This ambiguity leaves room for qualitatively new phenomena, and even new phases of matter, to arise.    A recent proposal of this kind is quarkyonic matter \cite{quarkyonic,spiral1,quarkyonic2,quarkyonic3,quarkyonic4,quarkyonic5,quarkyonic6,quarkyonic7}.
It is motivated by the ambiguity of defining confinement in a system where baryon density $\rho_B$ is high enough that there is $\sim \order{1}$ baryon per baryonic volume.

The possibility of quarkyonic matter \cite{quarkyonic} comes from the asymmetry between the confinement scale in temperature and chemical potential:
The amplitude of a gluon loop at finite temperature $ \sim N_c^2$, while a quark-hole loop at
finite chemical potential $\mu_q$  has amplitude $\sim N_f N_c \mu_q^2$ \cite{quarkyonic}.   While at high temperature ($T \geq \lqcd \sim N_c^0$)
confinement is broken by gluon loops alone, because of asymptotic freedom, at low temperature quark-hole loops need to overpower gluon loops.  
This requires $\mu_q \sim \sqrt{N_f/N_c} \lqcd$ at one loop, and an even higher exponent $z$ ($1/2<z <1$) at more than one loop \cite{quark1}.
As the number of colors might be considered ``large", this introduces an extra scale $\sqrt{N_c}\lqcd$ in momentum space relevant for deconfinement at finite chemical potential.   In configuration space, however, the only relevant scale is the inter-particle distance, which for one baryon per baryonic volume $\mu_q \sim \lqcd$ is always $\sim N_c^{-1/3} \rightarrow 0$.    

Dense matter at $\lqcd<\mu_Q<\sqrt{N_c/N_f}\lqcd$, with features of asymptotic freedom in configuration space
but features of confinement in momentum space, at $\lqcd<\mu_q< \sqrt{N_c/N_f} \lqcd$ is known as ``quarkyonic".
This is an interesting idea, but how much of quarkyonic dynamics survives at $N_c=3$ and $N_f=2,3$ is an open question.   It has long been known \cite{witten,manoharrev,quark4,quark5} that there are significant {\em qualitative} differences between the $N_c \rightarrow \infty$ limit and $N_c=3$.
As argued in \cite{quark1,quark3,quark4}, this indicates that the large $N_c$ limit is separated from the real world by a percolation-type phase transition. The quarkyonic matter transition line is therefore bound to be curved in $N_c$ as well as $T,\mu_B$ space, the former being accessible only on the lattice. 

The existence of quarkyonic matter, having the properties of \cite{quarkyonic} is therefore a matter for experimental investigation,  necessitating a quarkyonic matter {\em phenomenology}.     We shall attempt to develop one in this work, 
 based on the characteristics in \cite{quarkyonic} and their consequences explored in section 4 and 5 of \cite{quark1}.
Quark degrees of freedom make an appearance, and their interactions are governed by the Feynman rules of perturbative QCD.  The equation of state at equilibrium, therefore, is similar to that of an asymptotically free gas of quarks with a Fermi surface at $\mu_q \sim (1-3) \lqcd$ and low temperature.
Unlike ``real'' pQCD, however, confinement is still there: Baryons  continue to exist, and quark wavefunctions are localized around baryons.  As in the large $N_c$ limit, baryons are also approximately classical objects, well localized in position;
 They are also dense enough that there is, on average, one baryon per baryonic volume $\order{\lqcd^{-3}}$ or more.
  Hence, quark wavefunctions are {\em not} the asymptotically free quark wavefunctions of pQCD but are instead Eigenfunctions of a series of potential wells at the location of the baryons \cite{quark1} (Fig. \ref{diagfig}).
This is very similar to the dynamics of a free gas of electrons in a metal, where atoms are classical potential wells (as baryons are at large $N_c$ QCD) and electrons are fermions weakly interacting with each-other but with wavefunctions determined by classical potentials (as quarks are supposed to be in quarkyonic matter).
Summarizing, any dynamics inside quarkyonic matter will have pQCD interaction vertices, but incoming quark lines will pick up a form factor, reflecting their confinement.   Unlike in the vacuum, this form factor will not be ``localized'' (since percolation is naturally interpreted as the delocalization of quarks \cite{quark1}), but will reflect the dynamics of {\em all} baryons of the system \cite{quark1}.
At a single time step in configuration space the quark wavefunction looks like (arrows indicate a 3-vector, Greek indices a 4-vector)
\begin{equation}
\Psi(x) \propto  \sum_{i}^{hadrons} \phi\left(\vec{x}-\vec{x}_i\right) 
\end{equation}
 and $\phi(\vec{x}-\vec{x}_i)$ are peaks centered around the baryon location $x_i$ with wavefunction width in configuration space $\sim \lqcd^{-1}$, the confinement scale.  We approximate $\phi$ by Gaussian wavepackets
\begin{equation}
\phi(\vec{x}-\vec{x}_i)=\exp \left[ -\left(\vec{x}-\vec{x}_i \right)^2 \lqcd^2  \right].
\end{equation}
The advantage of this ansatz is that the mean field in a given event can be solved analytically:  The quark density in momentum space, assuming a baryon is a classical mean field of quarks, will be 
$\Psi^2(k)$, where
\begin{equation}
\label{psidef}
\tilde{\Psi}(k) \propto \sum_i \tilde{\phi} \left( \vec{k},\vec{x}_i \right)
\end{equation}
where  $\tilde{\phi}$ are the baryonic quark wavefunctions 
\begin{equation}
\tilde{\phi}(k,x_i) \propto \exp \left[ i \vec{k}\vec{x}_i - \vec{k}^2/\lqcd^2 \right]
\end{equation}
The space coordinates at each time-step $\vec{x}_i$ are extracted from a UrQMD \cite{Bass:1998ca,Bleicher:1999xi} simulation.   Note that the configuration space position of the baryon enters the wavefunction as a phase factor, to be multiplied with momentum.  The scattering rate will therefore pick up interference terms, a crucial effect for momentum anisotropy.
The important parameter in our calculation is the size of the baryon ``bag'' $\sim \lqcd^{-1}$, compared to the UrQMD-extracted distribution of baryons.   The ``bag size'' is relatively insensitive to in-medium modifications, in particular to a partial restoration of chiral symmetry (see discussion in \cite{quark1}).    Hence, in-medium modifications of vacuum hadron-hadron cross-sections, used in UrQMD should not impact the photon spectrum observables discussed here.
  We calculate Pb-Pb collisions at $\sqrt{s}=7.7$ GeV, within the FAIR energy range. We expect the $\sqrt{s}$ and system size dependence of them to be weak, allowing comparisons with any of \cite{low1,low2,low3,low4}.

What are the observable consequences of such dynamics?
Electromagnetic signals are sensitive to the earliest, densest phase.   Unlike, for example, hydrodynamic observables (that depend on the Equation of State), the form factors directly influence the final spectrum. Hence, the exploration of spectra of electromagnetic particles is an obvious place to distinguish between quarkyonic phases and more conventional Quark-Gluon Plasma (QGP). 
The first observable we look at is photon production from quark-quark pQCD scattering.   Photons (unlike, say, dileptons discussed in \cite{quark1}) are not so sensitive to parameters like the degree of thermalization of quarks and holes, about which little is known. 

To understand what process is most relevant for photon production in quarkyonic matter, we have to remember that quarks are delocalized by density effects, rather than deconfinement.    Hence, antiquarks and gluons remain localized \cite{quarkyonic,quarkyonic2,spiral1,ads}, and can be safely neglected within baryon structure functions at this $T,\mu_Q$.  
The $qg \rightarrow q \gamma,q \overline{q} \rightarrow g \gamma$ processes, 
dominating in a QGP \cite{muqpho1,muqpho2}, can be neglected, as can all processes with outgoing gluons and antiquarks.
The leading quark-level production process is then quark-quark Bremsstrahlung.
Its scattering matrix, studied in \cite{photons4}, is
 \begin{equation}
\mathcal{M}^2= L^2(k_1,k_2 \rightarrow k_3,k_4,p) + L^2(k_1 \leftrightarrow k_2,k_3 \leftrightarrow k_4 )
\label{meq}
\end{equation}
in terms of the fine structure constant $e$ and the QCD coupling constant $\lambda$. Here
$$ L^2 = -\frac{1}{4}e^2 \lambda^2 N_c^{-2} (k_2-k_4)^{-4}  Tr \left[ \slashed{ k}_4 \gamma^\sigma \slashed{k}_2 \gamma_\rho \right] Tr\left[ \slashed{k}_3 Z_\sigma^\mu  \slashed{k}_1 Z_\mu^\rho  \right]  $$
and
$Z_{\alpha}^{ \beta} =  \gamma_\alpha (k_1-p)^{-1} \gamma^\beta +\gamma^\beta (k_3+p)^{-1} \gamma_\alpha $.

  The photon rate convoluting the pQCD matrix and the wavefunction of quarkyonic matter is then
\begin{equation}
\label{spectrum}
\frac{dN_\gamma}{d^3 p} \propto \int \left( \mathcal{M} \tilde{\Psi}(k_1) \tilde{\Psi}^* (k_2)  \right)^2 d^3 k_{1,2,3,4} 
\end{equation}
where $\mathcal{M}$ is the matrix element corresponding to the diagrams Fig. \ref{diagfig} and Eq. \ref{meq}, production of a photon by the strong scattering of two quarks \cite{photons4} and $\tilde{\Psi}$ are given by Eq. \ref{psidef}.
Confinement, in both cases, is incorporated by removing quarks and gluons with momentum and virtuality $\leq \lqcd$.
\[\  \int d^3 k_{1,2,3,4} \rightarrow 
 \int_{\lqcd}^\infty k_{1,2,3,4}^2 dk_{1,2,3,4} \int d\Omega_{1,2,3,4}  \]
\[\ \times \Theta\left( (k_1+k_3)^2 - \lqcd^2 \right) \Theta\left( (k_2+k_4)^2 - \lqcd^2   \right) \]
which also takes care of collinear divergences.
\begin{figure}
\includegraphics[scale=0.15]{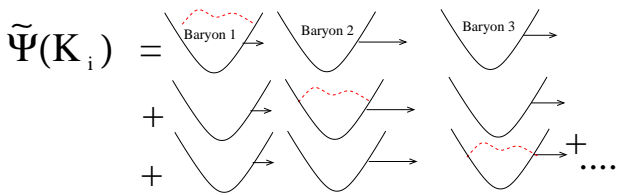}
\caption{(color online) The form of the quark wavefunction in quarkyonic matter, in 1D, as a red dashed wavy line.  Baryons are represented by semiclassical black potentials wells. Arrows depict the motion of the baryons. \label{potfig}}.
\end{figure}
 For the quarkyonic phase, Eq. \ref{spectrum} is calculated for each time step in each event, with the quark wavefunction reflecting baryon location for that event.  An average over UrQMD events is then obtained.  The integral in Eq. \ref{spectrum} was done by Monte Carlo, and the results, in particular the fluctuation in the last figure, were verified to be independent of statistics, both within and across events.   
No quark flow as separate from baryonic flow is included, as this would require separation of quarks from baryonic wave-functions.   Baryons, however, do develop collective flow, due to the comparatively strong baryon-baryon interactions, both due to scattering and mean field, UrQMD incorporates.   The backreaction of quarks to baryon flow is not understood (subleading in $N_c$), but the ``boosted quarkyonic'' scenario, described later, can be understood as an upper limit.

As a comparison, we also present the rate for an expanding thermalized quark-gluon plasma. As we do not consider the processes in \cite{muqpho1,muqpho2} this comparison is rough, but, because we are concentrating on spectra whose shapes are contained by local thermalization, the matrix element in Eq. \ref{meq} is sufficient for a qualitative estimate.
The only difference between a QGP and quarkyonic matter that the incoming quark distribution functions are boosted-thermal, with temperature $T$
\begin{equation}
\tilde{\Psi} (k) \tilde{\Psi}^*( k') \propto \delta \left( k'-k \right) \exp \left(- u_\mu k^\mu/T \right) . 
\end{equation}
Flow $u_\mu$ includes longitudinal expansion across the kinematic range parametrized by longitudinal flow rapidity $y_L$, and a transverse expansion $v_T(\phi)=v+ v_{2T}\cos(\phi)$ 
\begin{equation}
u^\mu = \frac{1}{\sqrt{1-\tanh(y_L)^2-v_T(\phi)^2}} \left( \begin{array}{c}  1 \\   v_T(\phi) \cos(\phi) \\   v_T(\phi) \sin(\phi)  \\ \tanh(y_L)  \end{array}   \right)
\end{equation}
We have  checked that the results presented later, in both the quarkyonic and thermal ansatze, are qualitatively similar if the QGP-appropriate $q g \rightarrow q\gamma,q \overline{q} \rightarrow g \gamma$ scattering processes are used in lieu of Bremstrahlung.   The results shown below are determined by the form of the wavefunction rather than the matrix element.

We now plot the transverse momentum ($p_T$) distributions as well as harmonic distributions w.r.t. the reaction plane $\Delta \phi = \phi-\phi_{RP}$.  The latter is parameterized with the $v_n$ coefficients, of which $v_2$ is the best-known example
\begin{equation}
\frac{dN_\gamma}{d^3 p} = \frac{dN_\gamma}{dy dp_T^2} \left( 1 + 2 \sum_{n=1}^\infty  v_n \left(p_T,y \right) \cos \left(\Delta \phi \right) \right) .
\end{equation}
\begin{figure}
\includegraphics[scale=0.22]{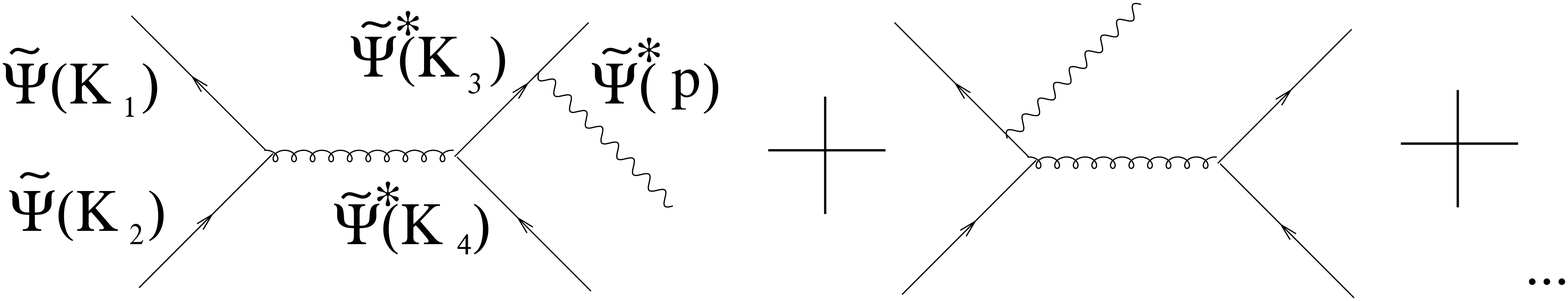}
\caption{(color online) The pQCD process we are examining \label{diagfig}. The emitted photon is denoted by a wavy line, the 
spring-line is a gluon while the solid lines are quarks.  The dominant process is $qq \rightarrow qq\gamma$ Brehmsstrahlung, which naturally leads for quarkyonic matter.  
Exchange diagrams can be obtained with the usual permutations}
\end{figure}
The discussion above makes it clear that quarkyonic matter and the most ``normal'' QGP can not, in general, coexist:  quarkyonic matter defined here is not simply a ``colder denser'' QGP, but a state where finite density-driven delocalization, which, due to the ``high'' baryon mass, is nearly vertical on the $T-\mu_B$ on the phase diagram (see Fig. 2 of \cite{quarkyonic})   Independently of initial chemical potential, cooling curves of the system created in heavy ion collisions are also steep on the $T-\mu_B$ axis \cite{petersen}. Therefore, quarkyonic matter should not appear in the cooling of a hotter QGP:   quarkyonic matter and QGP should appear as alternative scenarios, and should not coexist within the same event (In this sense, the constituent quark gas in \cite{csernai} is {\em not} quarkyonic matter).   

We also compare our model to calculations of direct photons from purely
hadronic UrQMD. For this, we employ the model developed
in~\cite{bjornurqmd}, in which hadronic scatterings from UrQMD are
considered as potential photon sources. The set of channels for photon
production and their differential cross-sections are taken from Kapusta
et al.~\cite{muqpho2}. They include scatterings of $\pi$, $\rho$
and $\eta$ mesons. Photons from scatterings at high momentum transfer
are neglected in calculations at FAIR energies. 

Missing in any $N_c \rightarrow \infty$ calculation are quark boosts due to finite baryon momentum, since baryons in this limit are static.
  Due to the delocalization of quarks, the consequences of finite momentum baryons are actually not so trivial, but they will always come with a $N_c^{-1}$ factor, consistent with the hierarchy between ``light'' quarks and heavy baryons.
In a $N_c=3$ world, corrections to this might be significant.
To estimate qualitatively the effect of these corrections, we choose a baryon at random in UrQMD, and ``localize" the quark to that baryon, Lorentz-boosting the quark by the baryon's momentum.   Since baryons have some flow on average, this boosts the flow of the quarks. This distinguishes ``quarkyonic'' from ``boosted quarkyonic'' in the plots.   While this is not a quantitative estimate, it is an ``extreme scenario'', where a delocalized quark receives a ``full boost'' from one particular baryon.   Hence, in a sense, it provides an upper limit to how large the baryonic flow contribution can be without quarks becoming the actively flowing degrees of freedom.

\begin{figure}
\vspace{0.5cm}
\includegraphics[scale=0.3]{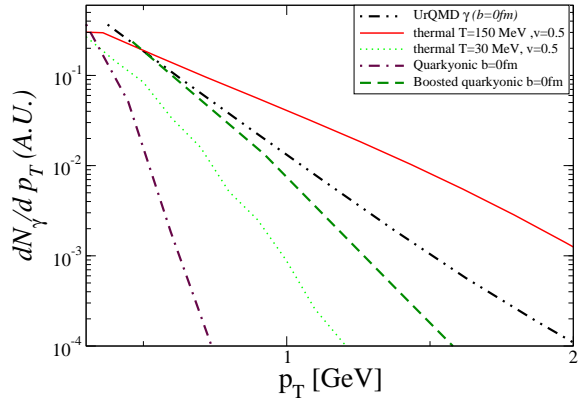}
\caption{(color online) The photon transverse momentum distribution for quarkyonic matter and thermalized QGP \label{ptfig}}
\end{figure}

\begin{figure}
\includegraphics[scale=0.3]{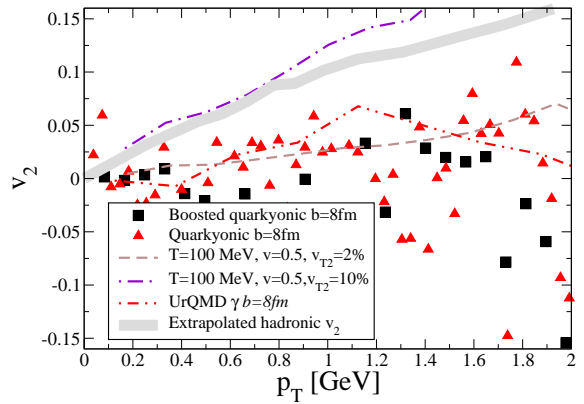}
\caption{The photon $v_2$ for quarkyonic matter and thermalized QGP \label{v2fig}.  See text for an explanation of the main result, the $v_2$ spread, which is independent of binning and statistics. An extrapolation from hadron $v_2$ is also shown}
\end{figure}

Similarly, this model is non-causal, 
since the quark wavefunctions adjust to baryon movement instantaneously.  
This is another artifact of the approximation discussed above, fixed by 
$\sim N_c^{-1}$ contributions, but {\em not} improved by the localization 
ansatz.   Improving on this approximation would mean making quarks off-shell, in a way that impacts baryon dynamics (or backward-in time signal propagation would occur).
For longitudinal dynamics, where typical baryon 
longitudinal rapidity $y_L \sim 1$, 
this could be a significant issue, but for transverse dynamics, where 
baryon speeds $y_T \ll 1$, this can safely be ignored since quarks are 
much faster than baryons.  Our most interesting results are indeed 
transverse. The results are shown in Fig. \ref{ptfig} for transverse momentum distribution and in Fig. \ref{v2fig} for the elliptic flow of Bremsstrahlung direct photons, at impact parameter $b=0,8$ fm (rapidity distributions are quantitatively similar between the three models, and qualitatively match experimental data).   
Normalization is arbitrary, as it is highly dependent on the undetermined strong coupling constant in the quark-quark scattering processes.   Any determination of quarkyonic matter would come from the {\em shape} of the distributions.


Fig. \ref{ptfig} shows the $p_T$ distribution for quarkyonic matter is distinctively steeper than realistic thermal curves, or the UrQMD analysis.   To reproduce it with a thermal curve, an unrealistically low mean temperature has to be used, one well below deconfinement, even at realistically large chemical potential.
A steep $p_T$ distribution, therefore, can be considered as a possible signature for quarkyonic matter.

We note that this steepness is natural to explain within quarkyonic assumptions:  quarks are delocalized, and hence do not feel the flow of any particular hadron.   
   Boosted UrQMD, unsurprisingly, is much less steep, but still well on the low side for temperature ($T \sim 30 $MeV), less than both hadronic and partonic dynamics \cite{urqmdphot,gale,gale2,huovinen,heinz}.

While photon elliptic flow, and spectra in general, have yet to be 
measured for the energies discussed here, previous experience \cite{lhcp,rhicp} 
suggests photon flow follows hadron flow closely. The latter, to a good 
approximation, is energy-independent when binned with $p_T$ 
\cite{starpaper}.  While this is something hydrodynamic and 
transport models have yet to 
account properly \cite{starpaper,scaling}, we can use it to extrapolate (thick line in Fig. \ref{v2fig}).

While the flowing ansatz reproduces the observed trends with the right choice of parameters, quarkyonic $v_2$ is something qualitatively different:  Overall $v_2$ is compatible with zero, but with strong variations {\em both} event by event {\em and} within the same event in different $p_T$ bins.   This spread is approximately constant with centrality.
This, while completely different from anything seen before, is actually physically not surprising:
If baryons are not moving, the 
only source of $v_2$ are the effects of the baryon distribution on the quark wavefunction.  The latter oscillate 
with a frequency $\sim p_T \rho_B^{-1/3}$ in an event-by-event dependent manner, determined by both the density and the quark momentum.     The inter-baryon distance $\rho_B^{-1/3}$ is highly inhomogeneus, both within the same event and event by event.   Hence, both that $v_2=0$ overall, and its random oscillation are not surprising.

For the experimental observability of this pattern, however, one must keep into account that it is generated by {\em wavefunctions}.  Fig. \ref{v2fig} is calculated in the limit of ``many photons per event'' as well as ``many events'', since it is in this limit that correlations due to the wavefunction shape become observable.   The oscillation amplitude and frequency, given ``many photons per event'', will {\em not} depend on the number of events, but only on $\rho_B^{-1/3}$ via the phase in Eq. \ref{psidef}, given by UrQMD.      In this limit, with $p_T$ bins narrower than $\rho_B^{1/3}$, the photon event-by-event $v_2$ will vary randomly with the amplitude and frequency given by Fig. \ref{v2fig}.   Away from this limit ($\sim \order{1}$ direct photon/event), a zero $v_2$ at all $p_T$ will be observed.  

Since this result is unaffected by baryonic flow ("Boosted Quarkyonic" also has average zero $v_2$), however, collapse of average $v_2$ can also be used as an indication of the quarkyonic phase, since photon $v_2$ is present in both QGP and hadrons (where anisotropies are dominated by flow rather than wavefunction shape).
As the UrQMD simulation in Fig. \ref{v2fig} shows, 
hadronic admixture, in particular photon decays from final-state 
hadrons, does not alter this conclusion because of the low $v_2(p_T)$ of direct photons in a hadronic medium.

     If $> \order{1}$ direct photons per event are detected and average $v_2$ is negligible, it is possible to look for the random fluctuations we describe, since $v_2$  oscillates {\em by momentum bin} and not just {\em by event}:      
A minimum of $\gamma$ $v_2$ in a $\sqrt{s}$ scan, associated with an increase of $v_2$ fluctuations separated from reaction-plane correlated charged particle harmonics, might be a signal for the onset of a quarkyonic-dominated regime.   \cite{newmethod} discusses ways to perform such an analysis.

In conclusion, we have calculated the $p_T$ and harmonic distribution of photons from quarkyonic matter, defined as a gas of perturbative quarks moving in a baryon generated
 classical potential.   We found that the photon spectrum is steeper than that expected from a QGP or a hadron gas \cite{urqmdphot} at similar temperature, and $v_2$ oscillates in a way different from both partonic and hadronic regimes. This difference can be understood from the shape of quark wavefunctions.   Provided the quarkyonic regime dominates over QGP  and the fluctuating component can be isolated, this effect can be developed into an experimental signature of quarkyonic matter.

We acknowledge the financial support received from the Helmholtz International
Centre for FAIR within the framework of the LOEWE program
(Landesoffensive zur Entwicklung Wissenschaftlich-\"Okonomischer
Exzellenz) launched by the State of Hesse.
This research used resources of the Oak Ridge Leadership Computing Facility at the Oak Ridge National Laboratory, which is supported by the Office of Science of the U.S. Department of Energy under Contract No. DE-AC05-00OR22725.
GT also acknowledges support from DOE under Grant No. DE-FG02-93ER40764. We thank the organizers of the FAIRNESS meeting, where the idea for of this work was initially developed, and Olena Linnyk and Mauricio Martinez Guerrero for discussions and suggestions.

\end{document}